\documentclass[12pt]{article}
\usepackage[dvips]{graphicx}

\begin{document}

\title{A Generalization of the Stillinger-Lovett Sum Rules
for the Two-Dimensional Jellium}

\author{L. {\v S}amaj$^1$}

\maketitle

\begin{abstract}
In the equilibrium statistical mechanics of classical Coulomb fluids,
the long-range tail of the Coulomb potential gives rise to the
Stillinger-Lovett sum rules for the charge correlation functions.
For the jellium model of mobile particles of charge $q$
immersed in a neutralizing background, the fixing of one of 
the $q$-charges induces a screening cloud of the charge density whose zeroth 
and second moments are determined just by the Stillinger-Lovett sum rules.
In this paper, we generalize these sum rules to the screening cloud
induced around a pointlike guest charge $Z q$ immersed in the bulk
interior of the 2D jellium with the coupling constant $\Gamma=\beta q^2$
($\beta$ is the inverse temperature), in the whole region of the
thermodynamic stability of the guest charge $Z>-2/\Gamma$.
The derivation is based on a mapping technique of the 2D jellium
at the coupling $\Gamma$ = (even positive integer) onto a discrete
1D anticommuting-field theory; we assume that the final results
remain valid for all real values of $\Gamma$ corresponding to
the fluid regime.
The generalized sum rules reproduce for arbitrary coupling $\Gamma$
the standard $Z=1$ and the trivial $Z=0$ results.
They are also checked in the Debye-H\"uckel limit $\Gamma\to 0$ and
at the free-fermion point $\Gamma=2$.
The generalized second-moment sum rule provides some exact information 
about possible sign oscillations of the induced charge density in space.
\end{abstract}

\medskip


\vfill

\noindent $^1$ 
Institute of Physics, Slovak Academy of Sciences, D\'ubravsk\'a cesta 9, 
\newline 845 11 Bratislava, Slovak Republic; e-mail: fyzimaes@savba.sk

\newpage

\renewcommand{\theequation}{1.\arabic{equation}}
\setcounter{equation}{0}

\section{Introduction}
The present paper deals with the equilibrium statistical mechanics
of a classical (i.e. non-quantum) jellium, sometimes called 
the one-component plasma, formulated in two spatial dimensions (2D).

The jellium model consists of mobile pointlike particles
$j=1,\ldots,N$ of charge $q$ and position vectors ${\bf r}_j$,
confined to a continuous domain $\Lambda$.
The particles are embedded in a spatially uniform neutralizing
background of charge density $-q n$.
The bulk regime of interest corresponds to the limits $N\to\infty$ and 
$\vert\Lambda\vert\to\infty$ with the fixed particle density 
$n=N/\vert\Lambda\vert$.

According to the laws of 2D electrostatics, the particles can be
thought of as infinitely long charged lines in the 3D space 
which are perpendicular to the confining 2D surface $\Lambda$.
Thus, the electrostatic potential $\phi$ at a point ${\bf r}\in\Lambda$,
induced by a unit charge at the origin ${\bf 0}$, is given by the 2D
Poisson equation
\begin{equation} \label{1.1}
\Delta \phi({\bf r}) = - 2\pi \delta({\bf r}) .
\end{equation}
For an infinite plane $\Lambda=R^2$, the solution of this equation,
subject to the boundary condition $\nabla \phi({\bf r})\to 0$ as
$\vert {\bf r}\vert\to\infty$, reads
\begin{equation} \label{1.2}
\phi({\bf r}) = - \ln \left( \frac{r}{r_0} \right) ,
\end{equation}
where $r\equiv \vert {\bf r}\vert$ and the free length constant $r_0$ 
will be set for simplicity to unity.
In the 2D Fourier space defined by
\begin{eqnarray} 
f(r) & = & \int \frac{{\rm d}^2 k}{2\pi}
\hat{f}(k) \exp({\rm i}{\bf k}\cdot {\bf r}) , \label{1.3} \\
\hat{f}(k) & = & \int \frac{{\rm d}^2 r}{2\pi}
f(r) \exp(-{\rm i}{\bf k}\cdot {\bf r}) \nonumber \\
& = & \sum_{j=0}^{\infty} \frac{(-1)^j}{(j!)^2} \left( \frac{k^2}{4} \right)^j
\frac{1}{2\pi} \int {\rm d}^2 r\, r^{2j} f({\bf r}) , \label{1.4}
\end{eqnarray}
the Coulomb potential (\ref{1.2}) exhibits the form 
\begin{equation} \label{1.5}
\hat{\phi}(k) = \frac{1}{k^2}
\end{equation}
with the characteristic singularity at $k=0$.
This maintain many generic properties of ``real'' 3D Coulomb fluids
with the interaction potential $\phi(r)=1/r$, ${\bf r}\in R^3$.

Because of the presence of the rigid background, the equilibrium
statistics of the jellium is usually studied in the canonical ensemble
under the condition of the overall charge neutrality.
The 2D statistics depends on the coupling constant $\Gamma=\beta q^2$
with $\beta=1/(k_{\rm B}T)$ being the inverse temperature;
the particle density $n$ only scales appropriately the distance.
Let the symbol $\langle\cdots\rangle_{\beta}$ denotes the canonical averaging.
At the one-particle level, one introduces the average number density
of particles
\begin{equation} \label{1.6}
n({\bf r}) = \Bigg\langle \sum_j \delta({\bf r}-{\bf r}_j) 
\Bigg\rangle_{\beta} .
\end{equation} 
At the two-particle level, one introduces the two-body density
\begin{equation} \label{1.7}
n^{(2)}({\bf r},{\bf r}') 
= \Bigg\langle \sum_{j\ne k} 
\delta({\bf r}-{\bf r}_j) \delta({\bf r}'-{\bf r}_k) 
\Bigg\rangle_{\beta} .
\end{equation} 
It is also useful to consider the pair correlation function
\begin{equation} \label{1.8}
h({\bf r},{\bf r}') =
\frac{n^{(2)}({\bf r},{\bf r}')}{n({\bf r}) n({\bf r}')} - 1 ,
\end{equation} 
which tends to $0$ at asymptotically large distances
$\vert {\bf r}-{\bf r}'\vert\to\infty$. 

The bulk jellium is in a fluid state for high enough temperatures, 
i.e. the density of particles is homogeneous, $n({\bf r})=n$, 
and the two-body density is translation invariant,
$n^{(2)}({\bf r},{\bf r}') = n^{(2)}(\vert {\bf r}-{\bf r}'\vert)$.
There are indications from numerical simulations \cite{Choquard83}
that around $\Gamma\sim 142$ the fluid system undergoes a phase
transition to a 2D Wigner crystal. 
In what follows, we shall restrict ourselves to the fluid region
of $\Gamma$-values.

Through a simple scaling argument, the exact equation of state for 
the pressure $P$, $\beta P = n[1-(\Gamma/4)]$, has been known for
long time \cite{Salzberg}.
The jellium is completely solvable, like any Coulomb system,
in the high-temperature Debye-H\"uckel (DH) limit $\Gamma\to 0$ \cite{Debye},
characterized by a monotonic exponential decay of the pair correlation
function $h(r)$ at asymptotically large distances $r\to\infty$.
The systematic $\Gamma$-expansion of statistical quantities around
the DH limit can be done within a bond-renormalized Mayer diagrammatic 
expansion \cite{Deutsch}. 
The 2D jellium is mappable onto a system of free fermions at the special
coupling $\Gamma=2$ \cite{Jancovici81}.
This exactly solvable point is characterized by a pure Gaussian decay of 
the pair correlation.
The evaluation of the leading term of the $(\Gamma-2)$ expansion
for $h(r)$ indicates the change from the monotonic to oscillatory behavior
just at $\Gamma=2$ \cite{Jancovici81}. 

The long-range tail of the Coulomb potential, which is reflected
in the singular behavior of the Fourier component (\ref{1.5}) around
$k=0$, causes screening and thus gives rise to exact constraints
(sum rules) for the charge correlation functions (see review \cite{Martin}),
like the zeroth- and second-moment Stillinger-Lovett conditions
\cite{Stillinger1,Stillinger2}.
Their derivation can be based on the exploration of 
the Ornstein-Zernicke (OZ) equation
\begin{equation} \label{1.9}
h({\bf r},{\bf r}') = c({\bf r},{\bf r}') + \int {\rm d}^2 r''\,
c({\bf r},{\bf r}'') n({\bf r}'') h({\bf r}'',{\bf r}') 
\end{equation}
relating the pair correlation function $h$ with the {\em direct}
correlation function $c$.
Within the diagrammatic scheme of the renormalized Mayer expansion
\cite{Deutsch}, the direct correlation function of the bulk jellium
is expressible as
\begin{equation} \label{1.10}
c(r) = - \beta q^2 \phi(r) + c_{\rm reg}(r) ,
\end{equation}
where $c_{\rm reg}$ denotes contributions of all completely
renormalized Mayer diagrams.
Since these contributions are short-ranged, the Fourier transform
of $c_{\rm reg}$ has an analytic $k$-expansion around $k=0$.
Consequently, as $k\to 0$,
\begin{equation} \label{1.11}
\hat{c}(k) = - \frac{\Gamma}{k^2} + O(1) .
\end{equation}
Writing the OZ equation (\ref{1.9}) in the 2D Fourier space
\begin{equation} \label{1.12}
\hat{h}(k) = \hat{c}(k) + 2\pi n \hat{c}(k) \hat{h}(k) ,
\end{equation}
the small-$k$ expansion of $\hat{c}$ (\ref{1.11}) fixes
the zeroth and second moments of $h(r)$.
In terms of the two-body density, these sum rules read
\begin{eqnarray}
\int {\rm d}^2 r \left[ n^{(2)}({\bf r},{\bf 0}) - n^2 \right]
& = & - n , \label{1.13} \\
\int {\rm d}^2 r \vert {\bf r} \vert^2
\left[ n^{(2)}({\bf r},{\bf 0}) - n^2 \right]
& = & - \frac{2}{\pi\Gamma} . \label{1.14}
\end{eqnarray}
It is clear from the derivation procedure that the consideration
of a short-ranged, e.g. hard core, potential in addition to the Coulomb 
potential does not alter the results (\ref{1.13}) and (\ref{1.14}).
We add for completeness that for the 2D jellium also the fourth-moment
condition \cite{Vieillefosse} (related to the availability of 
the exact equation of state) and the sixth-moment condition \cite{Kalinay}
(derived within a classification of renormalized Mayer diagrams)
are known. 

In this paper, we study a typical situation in the theory of 
colloidal mixtures \cite{Levin,Tellez06}: a ``guest'' particle with charge 
$Z q$ is immersed into the bulk interior of a Coulomb system,
in our case the jellium.
Possible values of the parameter $Z$ are restricted as follows.
When $q$ is the elementary charge $e$ of an electron, $Z$ is the valence 
of an atom and as such it should be an integer.
In general, the jellium can be composed of multivalent charges 
$(q=\pm 2e,\pm 3e,\ldots)$ and in that case $Z$ can take rational values.
In the considered case of the {\em pointlike} guest charge and two
spatial dimensions, the value of $Z$ is bounded from below 
by a collapse phenomenon.
Namely, the Boltzmann factor of the guest charge $Z q$ with 
a jellium charge $q$ at distance $r$, $r^{\Gamma Z}$, is integrable 
at small 2D distances $r$ if and only if 
\begin{equation} \label{1.15}
Z > - \frac{2}{\Gamma} .
\end{equation}
This is the region of the thermodynamic stability for the jellium
system plus the guest charge $Zq$.  

The aim of the present paper is to extend the Stillinger-Lovett 
sum rules (\ref{1.13}) and (\ref{1.14}) to the presence of the guest 
charge $Z q$ in the bulk jellium.
For this purpose, we introduce ``conditional'' densities: 
let $n({\bf r}\vert Zq,{\bf 0})$ be the average density of jellium
particles at point ${\bf r}$ induced by a pointlike charge $Z q$
placed at the origin ${\bf 0}$.
The corresponding induced charge density will be denoted by
$\rho({\bf r}\vert Zq,{\bf 0}) = q [ n({\bf r}\vert Zq,{\bf 0}) - n]$. 
Evidently, if $Z=1$, i.e. the fixed particle has the same charge 
as the species forming the jellium, it holds
\begin{equation} \label{1.16}
n^{(2)}({\bf r},{\bf 0}) = n({\bf r}\vert q,{\bf 0}) n({\bf 0}) .  
\end{equation}
The sum rules (\ref{1.13}) and (\ref{1.14}) can be thus rewritten
in the form
\begin{eqnarray}
\int {\rm d}^2 r \rho({\bf r}\vert q,{\bf 0}) & = & - q , \label{1.17} \\
\int {\rm d}^2 r \vert {\bf r} \vert^2 \rho({\bf r}\vert q,{\bf 0})
& = & - \frac{2 q}{\pi\Gamma n} . \label{1.18}
\end{eqnarray}
The zeroth-moment condition (\ref{1.17}) reflects a trivial fact
that the charge $q$ is screened by a cloud of the opposite charge $-q$.
The condition (\ref{1.18}) tells us that the second-moment of this
charge cloud has a prescribed value.
Our task is to generalize these exact constraints for the conditional
charge density $\rho({\bf r}\vert Zq,{\bf 0})$, where the guest-charge
parameter $Z$ lies in the stability region (\ref{1.15}).
We notice that there exists one trivial case $Z=0$, for which
the obvious equality $n({\bf r}\vert 0,{\bf 0}) = n$ implies
that all charge moments vanish,
\begin{equation} \label{1.19}
\int {\rm d}^2 r \vert {\bf r} \vert^{2j}
\rho({\bf r}\vert 0,{\bf 0}) = 0 \quad
\mbox{for $j=0,1,2,\ldots$.} 
\end{equation}

The generalization of the zeroth-moment relation (\ref{1.17}) is 
straightforward:
\begin{equation} \label{1.20}
\int {\rm d}^2 r \rho({\bf r}\vert Zq,{\bf 0}) = - Zq ,
\end{equation}
i.e., the guest charge $Z q$ immersed in the jellium is screened 
by an excess cloud of jellium particles carrying exactly the 
opposite charge $-Z q$.

The generalization of the second-moment relation (\ref{1.18}) is nontrivial.
We would like to emphasize that the derivation of the sum rule
(\ref{1.14}), or its equivalent (\ref{1.18}), using the OZ equation
was based on the translation-invariance property of the bulk jellium.
The introduction of the guest charge $Z q$ with $Z\ne 1$ breaks 
the translation symmetry of the jellium and one has therefore to apply 
other more sophisticated approaches.
Here, we use a mapping technique of the 2D jellium with the coupling 
constant $\Gamma$ = (even positive integer) onto a discrete 1D 
anticommuting-field (fermion) theory, introduced in ref. \cite{Samaj95} and
developed further in refs. \cite{Samaj98,Samaj00,Samaj04}.
The general formalism of the mapping technique is briefly recapitulated
in Section 2.
 
The present application of the fermionic mapping to the thermodynamic limit 
of the jellium in the disc geometry, with the guest charge $Z q$ fixed at 
the disc center, is the subject of Section 3.
Within the fermion representation, a couple of constraints for fermionic
correlators is derived by using specific transformations of 
anticommuting variables.  
Under the assumption of good screening properties of the jellium system,
these fermionic constraints imply the electroneutrality sum rule
(\ref{1.20}) and the desired second-moment sum rule:
\begin{equation} \label{1.21}
\int {\rm d}^2 r \vert {\bf r} \vert^2
\rho({\bf r}\vert Zq,{\bf 0}) = - \frac{1}{\pi\Gamma n} Zq
\left[ \left( 2-\frac{\Gamma}{2} \right) + \frac{\Gamma}{2} Z \right] , 
\end{equation}
valid in the guest-charge stability region (\ref{1.15}).
Although this relation was obtained for the series of
discrete values of the coupling constant $\Gamma = 2,4,\ldots$,
we assume its validity for all real values of $\Gamma$ corresponding to
the fluid regime.
It is easy to verify that the formula (\ref{1.21}) is consistent for $Z=1$
with the result (\ref{1.18}) and for $Z=0$ with Eq. (\ref{1.19}).
In contrast to the zeroth-moment condition (\ref{1.20}), the second-moment
sum rule (\ref{1.21}) provides some exact information about
possible sign oscillations of the charge cloud screening the guest 
particle $Z q$ and this topic is also discussed in Section 3.

The exact weak-coupling DH analysis of the studied guest-charge problem 
is presented in Section 4, with the final result
\begin{equation} \label{1.22}
\int {\rm d}^2 r \vert {\bf r} \vert^2
\rho({\bf r}\vert Zq,{\bf 0}) = - \frac{2 Zq}{\pi\Gamma n}
\qquad \mbox{as $\Gamma\to 0$.}
\end{equation}
The crucial formula (\ref{1.21}) evidently passes this test.

The exact treatment of the problem at the free fermion point $\Gamma=2$,
performed in Section 5, leads for stable $Z>-1$ to the result
\begin{equation} \label{1.23}
\int {\rm d}^2 r \vert {\bf r} \vert^2
\rho({\bf r}\vert Zq,{\bf 0}) = - \frac{Zq(Z+1)}{2\pi n}
\qquad \mbox{at $\Gamma=2$.}
\end{equation}
The formula (\ref{1.21}) passes also this test.

Some concluding remarks are given in Section 6.

\renewcommand{\theequation}{2.\arabic{equation}}
\setcounter{equation}{0}

\section{General formalism}
Let us consider the jellium consisting of $N$ mobile particles
confined to a 2D domain $\Lambda$; the plain hard walls surrounding
$\Lambda$ do not produce image charges.
In terms of the complex coordinates $(z,\bar{z})$,
the potential energy of the particle-background system is given by
\begin{equation} \label{2.1}
E = E_0 + q\sum_j \phi(z_j,\bar{z}_j) -
q^2 \sum_{j<k} \ln \vert z_j-z_k \vert .
\end{equation}
Here, $\phi(z,\bar{z})$ is the one-body potential induced by
the background plus perhaps some additional fixed charges and
$E_0$ is the (background-background, etc.) interaction constant 
which does not influence the statistical averages over particle
positions and therefore will be omitted.
The canonical partition function at the inverse temperature $\beta$
reads
\begin{equation} \label{2.2}
Z_N = \frac{1}{N!} \int_{\Lambda} \prod_{j=1}^N
\left[ {\rm d}^2 z_j w(z_j,\bar{z}_j) \right]
\prod_{j<k} \vert z_j - z_k \vert^{\Gamma} ,
\end{equation}
where the one-body Boltzmann factor 
$w(z_j,\bar{z}_j) = \exp[-\beta q \phi(z_j,\bar{z}_j)]$.
The particle density (\ref{1.6}) can be obtained in the standard way
\begin{equation} \label{2.3}
n(z,\bar{z}) = w(z,\bar{z}) \frac{\delta \ln Z_N}{\delta w(z,\bar{z})} .
\end{equation}

For the coupling constant $\Gamma=2\gamma$ ($\gamma=1,2,\ldots$ an integer),
it has been shown in ref. \cite{Samaj95} that the partition function 
(\ref{2.2}) can be expressed as the integral over
two sets of Grassmann variables $\{ \xi_j^{(\alpha)},\psi_j^{(\alpha)}\}$
each with $\gamma$ components ($\alpha = 1,\ldots,\gamma$), defined
on a discrete chain of $N$ sites $j=0,1,\ldots,N-1$ and satisfying 
the ordinary anticommuting algebra \cite{Berezin}, as follows:
\begin{eqnarray}
Z_N & = & \int {\cal D}\psi {\cal D}\xi \exp \left[ S(\xi,\psi) \right] , 
\label{2.4} \\
S(\xi,\psi) & = &  \sum_{j,k=0}^{\gamma(N-1)} \Xi_j w_{jk} \Psi_k . 
\label{2.5}
\end{eqnarray}
Here, ${\cal D}\psi {\cal D}\xi = \prod_{j=0}^{N-1}
{\rm d}\psi_j^{(\gamma)} \ldots {\rm d}\psi_j^{(1)}
{\rm d}\xi_j^{(\gamma)} \ldots {\rm d}\xi_j^{(1)}$ 
and the action $S$ involves pair interactions of ``composite'' operators
\begin{equation} \label{2.6}
\Xi_j = \sum_{j_1,\ldots,j_{\gamma}=0\atop (j_1+\cdots +j_{\gamma})=j}^{N-1} 
\xi_{j_1}^{(1)} \cdots \xi_{j_{\gamma}}^{(\gamma)} , \qquad
\Psi_k = \sum_{k_1,\ldots,k_{\gamma}=0\atop (k_1+\cdots +k_{\gamma})=k}^{N-1} 
\psi_{k_1}^{(1)} \cdots \psi_{k_{\gamma}}^{(\gamma)} .
\end{equation} 
The interaction strength is given by
\begin{equation} \label{2.7}
w_{jk} = \int_{\Lambda} {\rm d}^2 z\, w(z,\bar{z}) z^j \bar{z}^k ;
\qquad j,k = 0,1,\ldots,\gamma(N-1) .
\end{equation}
Using the notation 
$\langle \cdots \rangle = \int {\cal D}\psi {\cal D}\xi {\rm e}^S \cdots/Z_N$ 
for an averaging over the anticommuting variables with the action (\ref{2.5}), 
the particle density (\ref{2.3}) is expressible in the fermionic format 
as follows
\begin{equation} \label{2.8}
n(z,\bar{z}) = w(z,\bar{z}) \sum_{j,k=0}^{\gamma(N-1)}
\langle \Xi_j \Psi_k \rangle z^j \bar{z}^k .
\end{equation}

Specific constraints for the fermionic correlators 
$\langle \Xi_j\Psi_k \rangle$ follow from the fermionic representation of 
the partition function as the results of certain transformations of 
anticommuting variables which maintain the composite nature of 
the action (\ref{2.5}).

Let us first rescale by a constant one of the field components, say
\begin{equation} \label{2.9}
\xi_j^{(1)} \to \mu \xi_j^{(1)}  \qquad j=0,1,\ldots, N-1 . 
\end{equation}
Jacobian of this transformation equals to $\mu^N$ and the fermionic
action $S$ transforms to $\mu S$.
Consequently,
\begin{equation} \label{2.10}
Z_N = \mu^{-N} \int {\cal D}\psi {\cal D}\xi 
\exp\left( \mu \sum_{j,k=0}^{\gamma(N-1)}
\Xi_j w_{jk} \Psi_k \right) .
\end{equation}
$Z_N$ is independent of $\mu$ and so its derivative with respect
to $\mu$ is equal to zero for any value of $\mu$.
In the special case $\mu=1$, the equality
$\partial_{\mu} \ln Z_N \vert_{\mu=1} = 0$ implies the constraint
\begin{equation} \label{2.11}
\sum_{j,k=0}^{\gamma(N-1)} w_{jk} \langle \Xi_j \Psi_k \rangle = N .
\end{equation}
 
Let us now consider another linear transformation of {\em all}
$\xi$-field components
\begin{equation} \label{2.12}
\xi_j^{(\alpha)} \to \lambda^j \xi_j^{(\alpha)} \qquad
j=0,1,\ldots, N-1; \qquad \alpha = 1,\ldots,\gamma . 
\end{equation}
Jacobian of this transformation equals to $\lambda^{\gamma N(N-1)/2}$ and 
the fermionic action $S$ transforms to 
$\sum_{j,k=0}^{\gamma(N-1)} \lambda^j \Xi_j w_{jk} \Psi_k$.
Consequently,
\begin{equation} \label{2.13}
Z_N = \lambda^{-\gamma N(N-1)/2} \int {\cal D}\psi {\cal D}\xi 
\exp\left( \sum_{j,k=0}^{\gamma(N-1)} \lambda^j \Xi_j w_{jk} \Psi_k \right) .
\end{equation}
The equality $\partial_{\lambda} \ln Z_N \vert_{\lambda=1} = 0$ implies 
the following constraint
\begin{equation} \label{2.14}
\sum_{j,k=0}^{\gamma(N-1)} j w_{jk} \langle \Xi_j \Psi_k \rangle
= \frac{1}{2} \gamma N (N-1) .
\end{equation}
The application of the transformation (\ref{2.12}) to all 
$\psi$-field components leads to the complementary condition
\begin{equation} \label{2.15}
\sum_{j,k=0}^{\gamma(N-1)} k w_{jk} \langle \Xi_j \Psi_k \rangle
= \frac{1}{2} \gamma N (N-1) .
\end{equation}

\renewcommand{\theequation}{3.\arabic{equation}}
\setcounter{equation}{0}

\section{Derivation of sum rules}
We study the jellium model confined to the domain of disc geometry 
$\Lambda = \{ {\bf r},r<R \}$, with the guest charge $Zq$ fixed 
at the origin ${\bf 0}$.
The guest charge $Zq$ together with the total charge $Nq$ of $N$
mobile particles are compensated by the fixed background of charge
density $-nq$ via the overall neutrality condition
\begin{equation} \label{3.1}
Z + N = \pi R^2 n .
\end{equation} 
The potential induced by the homogeneous background is $q\pi nr^2/2$,
the guest charge interacts with jellium particles logarithmically $-Zq\ln r$.
The total one-body potential acting on each particle
\begin{equation} \label{3.2}
\phi({\bf r}) = q^2 \frac{\pi n r^2}{2} - Z q^2 \ln r  
\end{equation}
possesses the circular symmetry.

At the coupling $\Gamma=2\gamma$ ($\gamma=1,2,\ldots$), the one-body
Boltzmann factor $w({\bf r}) = \exp[-\beta\phi({\bf r})]$ reads
\begin{equation} \label{3.3}
w({\bf r}) = r^{2\gamma Z} \exp(-\gamma\pi n r^2) .
\end{equation}
Within the fermionic representation of the jellium (\ref{2.4})-(\ref{2.8}), 
the interaction matrix (\ref{2.7}) becomes diagonal
\begin{equation} \label{3.4}
w_{jk} = \delta_{jk} w_j , \qquad
w_j = \int_{\Lambda} {\rm d}^2 r\, r^{2(\gamma Z+j)} \exp(-\gamma\pi n r^2) . 
\end{equation}
The consequent diagonalization of the action (\ref{2.5}) in composite
operators, $S=\sum_{j=0}^{\gamma(N-1)} \Xi_j w_j \Psi_j$, implies that
$\langle \Xi_j\Psi_k \rangle = \delta_{jk} \langle \Xi_j\Psi_j \rangle$
and the representation of the particle density (\ref{2.8}) simplifies to
\begin{equation} \label{3.5}
n({\bf r}\vert Zq,{\bf 0}) = {\rm e}^{-\gamma \pi n r^2}
\sum_{j=0}^{\gamma(N-1)} \langle \Xi_j\Psi_j \rangle
r^{2(\gamma Z+j)} .
\end{equation}
The constraint (\ref{2.11}) is expressible as
\begin{equation} \label{3.6}
\sum_{j=0}^{\gamma (N-1)} w_j \langle \Xi_j\Psi_j \rangle = N
\end{equation}
and the couple of complementary conditions (\ref{2.14}) and (\ref{2.15})
reduces to
\begin{equation} \label{3.7}
\sum_{j=0}^{\gamma (N-1)} j w_j \langle \Xi_j\Psi_j \rangle = 
\frac{1}{2} \gamma N(N-1) .
\end{equation}

Using the definition of the interaction integrals (\ref{3.4}), 
it is easy to show that the constraint (\ref{3.6}) is equivalent to
the relation
\begin{equation} \label{3.8}
\int_{\Lambda} {\rm d}^2 r\, n(r\vert Zq,{\bf 0})  =  N ,
\end{equation}
which reflects a trivial fact: the total number of mobile particles
in the disc domain $\Lambda$ is equal to $N$.
With regard to the electroneutrality condition (\ref{3.1}),
the relation (\ref{3.8}) can be rewritten in the form
\begin{equation} \label{3.9}
\int_{\Lambda} {\rm d}^2 r \rho(r\vert Zq,{\bf 0}) = - Zq .
\end{equation}
By a simple analysis we shall argue that this condition involves
in fact two sum rules, the bulk one and the surface one.
Let us divide the disc domain $\Lambda$ onto its ``bulk'' part
$\Lambda_b = \{ {\bf r},r<R/2\}$ and the ``surface'' part
$\Lambda_s = \{ {\bf r},\mbox{$r=R-x$ with $0\le x<R/2$}\}$
($x$ denotes the distance from the disc boundary) 
and rewrite Eq. (\ref{3.9}) as follows 
\begin{equation} \label{3.10}
\int_0^{R/2} 2\pi r {\rm d}r \rho(r\vert Zq,{\bf 0})
+ \int_0^{R/2} 2\pi (R-x) {\rm d}x \rho(x\vert Zq,{\bf 0}) = - Zq .
\end{equation}
Let us assume that the system of charges has good screening properties,
i.e. the decay of particle correlations at large distances $r$ is faster
than any inverse power law, say exponential $\propto \exp(-\kappa r)$ 
with $\kappa$ being the inverse correlation length (like it is in 
the weak-coupling limit $\Gamma\to 0$) or even Gaussian 
$\propto \exp[-(\kappa r)^2]$ (like it is at the free-fermion point 
$\Gamma=2$). 
In the $R\to\infty$ limit, the particle density differs from 
the constant $n$ only: in the bulk region close to the disc center ${\bf 0}$ 
(up to $r\sim\kappa^{-1}$) and in the surface region close to 
the $x=0$ boundary (up to $x\sim\kappa^{-1}$).
The charge profile close to the boundary $\rho(x\vert Zq,{\bf 0})$ 
is influenced by the screened guest charge $Zq$ (exponentially or
even Gaussianly) weakly as $R\to\infty$. 
Forgetting these small terms, one can put
\begin{equation} \label{3.11}
\rho(x\vert Zq,{\bf 0}) \sim \rho(x\vert 0,{\bf 0}) = \rho(x)
+ \frac{1}{R} f_1(x) + \frac{1}{R^2} f_2(x) + \cdots ,
\end{equation}
where the long-ranged inverse-power-law terms $1/R, 1/R^2,\ldots$
are due to the nonzero curvature of the disc surface and the 
respective coefficients $f_1, f_2, \ldots$ are short-ranged
functions of the dimensionless parameter $\kappa x$.
Thus, Eq. (\ref{3.10}) splits in the limit $R\to\infty$ into
the $Z$-dependent bulk electroneutrality condition of present interest
\begin{equation} \label{3.12}
\int {\rm d}^2 r \rho(r\vert Zq,{\bf 0}) = - Zq  
\end{equation}
and a series of $Z$-independent surface conditions
\begin{equation} \label{3.13}
\int_{0}^{\infty} 2\pi (R-x) {\rm d}x \rho(x\vert 0,{\bf 0}) = 0 ,
\end{equation}
the lowest one of which takes the form of the surface electroneutrality
\begin{equation} \label{3.14}
\int_0^{\infty} {\rm d}x \rho(x) = 0 .
\end{equation}

To make use of the constraint (\ref{3.7}), we first differentiate
both sides of the density representation (\ref{3.5}) with respect to $r$,
then multiply the result by $r$ and finally integrate over the disc domain, 
to obtain
\begin{equation} \label{3.15}
\int_{\Lambda} {\rm d}^2 r\, r \frac{\partial}{\partial r}
n(r\vert Zq,{\bf 0}) = 2\gamma Z N 
- 2\gamma \pi n \int_{\Lambda} {\rm d}^2 r\, r^2 n(r\vert Zq,{\bf 0}) 
+ 2\sum_{j=0}^{\gamma(N-1)} j w_j \langle \Xi_j \Psi_j \rangle .
\end{equation}
The lhs of this relation can be integrated by parts, the summation
on the rhs is given by the constraint of interest (\ref{3.7}).
After simple algebra, the relation (\ref{3.15}) is transformed to
\begin{eqnarray} 
- 2\pi\gamma n \int_{\Lambda} {\rm d}^2 r\, r^2 
\rho(r\vert Zq,{\bf 0}) & = &
(2-\gamma) Zq + \gamma Z^2 q 
\nonumber \\ & & + 2\pi R^2  \left[ \rho(R\vert Zq,{\bf 0}) +
\frac{\gamma}{2} q n \right] . \label{3.16}
\end{eqnarray}
Like in the previous analysis of Eq. (\ref{3.9}), we divide the disc
domain $\Lambda$ onto its bulk and surface parts to express the integral in
Eq. (\ref{3.16}) as follows: 
\begin{equation} \label{3.17}
\int_0^{R/2} 2\pi r^3 {\rm d}r \rho(r\vert Zq,{\bf 0})
+ \int_0^{R/2} 2\pi (R-x)^3 {\rm d}x \rho(x\vert Zq,{\bf 0}) .
\end{equation}
Under the assumption of good screening properties of the jellium,
the bulk and surface regions are coupled weakly in the $R\to\infty$ limit 
and one can consider once more the expansion (\ref{3.11}) for 
the boundary charge density.
In this way, one gets from Eq. (\ref{3.16}) the $Z$-dependent bulk condition
\begin{equation} \label{3.18}
\int {\rm d}^2 r\, r^2
\rho(r\vert Zq,{\bf 0}) = - \frac{1}{2 \pi\gamma n} Zq
\left[ (2-\gamma) + \gamma Z \right] , 
\end{equation}
which is equivalent after the substitution $\gamma=\Gamma/2$ to 
the one of primary importance (\ref{1.21}),
and a series of $Z$-independent surface conditions
\begin{equation} \label{3.19}
-2\pi\gamma n \int_{0}^{\infty} 2\pi (R-x)^3 {\rm d}x \rho(x\vert 0,{\bf 0})
= 2\pi R^2 \left[ \rho(x=0\vert 0,{\bf 0}) + \frac{\gamma}{2} q n \right] .
\end{equation}
The lowest-order surface condition can be obtained by summing 
Eq. (\ref{3.13}), multiplied by $2\pi\gamma nR^2$, with Eq. (\ref{3.19}).
The final result reads
\begin{equation} \label{3.20}
\rho(x=0) = - \frac{\gamma}{2} q n
+ 4\pi \gamma n \int_0^{\infty} {\rm d}x\, x \rho(x) .
\end{equation}
This relation is known as the contact theorem 
\cite{Choquard80,Totsuji,Tellez99}.
Although all relations were derived for $\gamma=\Gamma/2$ a positive integer,
it is reasonable to extend their validity to all values of $\Gamma$
corresponding to the fluid regime.

As was mentioned in the Introduction, the generalized second-moment
sum rule (\ref{1.21}) is consistent with the available
results (\ref{1.18}) for $Z=1$ and (\ref{1.19}) for the trivial case $Z=0$. 
In the next two sections, we test this sum rule also in the weak-coupling
$\Gamma\to 0$ limit (Section 4) and at the free-fermion point $\Gamma=2$
(Section 5).

In contrast to the zeroth-moment electroneutrality condition (\ref{3.12}),
the generalized second-moment sum rule (\ref{3.18}), or equivalently
(\ref{1.21}), provides an exact information about possible sign
oscillations of the induced charge density $\rho(r\vert Zq,{\bf 0})$
in space.
If $Z>0$, the guest particle and jellium charges repeal each other
and therefore $\rho(r\vert Zq,{\bf 0})\sim -q n$ as $r\to 0$.
Provided that $\rho(r\vert Zq,{\bf 0})$ does not change the sign
when changing $r$ from $0$ to $\infty$ (where $\rho$ vanishes),
its second moment has the sign opposite to $Z q$.
Similarly, if $Z<0$, there is an attraction between the guest particle
and jellium charges, so that $\rho(r\vert Zq,{\bf 0})$ goes to infinity
as $r\to 0$. 
Consequently, when $\rho(r\vert Zq,{\bf 0})$ does not change the sign
when going from $r=0$ to $r\to\infty$, its second moment has again
the sign opposite to $Z q$.
The {\em sufficient} condition for sign oscillations 
of the charge density $\rho(r\vert Zq,{\bf 0})$ in space is 
that its second-moment has the sign of $Z q$. 
In view of the result (\ref{1.21}), the sufficient condition
for oscillations is that the guest-charge parameter $Z$ lies
in the interval
\begin{equation} \label{3.21}
- \frac{2}{\Gamma} < Z < 1 - \frac{4}{\Gamma} ,
\end{equation}
where the lower bound, see Eq. (\ref{1.15}), ensures the
thermodynamic stability of the pointlike guest charge $Z q$.
The inequalities (\ref{3.21}) have no solution for $\Gamma\le 2$.
For $\Gamma>4$, there exists also an interval of {\em positive} values of 
$Z$ for which the induced charge density certainly exhibits sign oscillations.

\renewcommand{\theequation}{4.\arabic{equation}}
\setcounter{equation}{0}

\section{Weak-coupling limit}
The effective potential $\phi$ at distance $r$ from the guest
charge $Z q$, placed at the origin ${\bf 0}$ and surrounded by mobile 
$q$-charges of the average density $n({\bf r}\vert Zq,{\bf 0})$ plus
the neutralizing background of charge density $-q n$, is given by
the 2D Poisson equation
\begin{equation} \label{4.1}
\Delta \phi({\bf r}) = - 2 \pi q \left\{ Z \delta({\bf r})
+ \left[ n({\bf r}\vert Zq,{\bf 0}) - n \right] \right\} .
\end{equation}

The weak-coupling (high-temperature) region $\Gamma\to 0$ is
described rigorously by the Debye-H\"uckel theory \cite{Debye,Kennedy}.
Within this mean-field approach, the average particle density
at a given point is approximated by replacing the potential of mean
force by the average electrostatic potential at that point,
$n({\bf r}\vert Zq,{\bf 0}) = n \exp[-\beta q \phi({\bf r})]$.
The mean-field Boltzmann factor can be linearized at high temperatures,
$\exp[-\beta q \phi({\bf r})] \sim 1 - \beta q \phi({\bf r})$.
The Poisson Eq. (\ref{4.1}) then reads
\begin{equation} \label{4.2}
\left( \Delta - \kappa^2 \right) \phi({\bf r}) = - 2 \pi Zq \delta({\bf r}) ,
\end{equation}
where $\kappa = \sqrt{2\pi\Gamma n}$ is the inverse Debye length. 

Due to the circular symmetry of the problem, 
$\Delta = \partial_r^2 + (1/r) \partial_r$.
Eq. (\ref{4.2}), subject to the condition of regularity at $r\to\infty$,
thus implies
\begin{equation} \label{4.3}
\phi({\bf r}) = Z q K_0(\kappa r) ,
\end{equation}
where $K_0$ is a modified Bessel function \cite{Gradshteyn}.

The induced charge density around the guest charge $Zq$ is obtained
in the form
\begin{equation} \label{4.4}
\rho({\bf r}\vert Zq,{\bf 0}) = - Zq n \Gamma K_0(\kappa r) .
\end{equation}
Since the stability lower bound (\ref{1.15}) is $Z>-\infty$
in the limit $\Gamma\to 0$, this result applies to all real values of $Z$.
The charge density (\ref{4.4}) is always a monotonic function of 
the distance $r$ which keeps its plus ($Zq<0$) or minus ($Zq>0$) sign 
in the whole interval of $r\in(0,\infty)$.
Its moments $I_j = \int_0^{\infty} 2\pi r {\rm d}r\, r^{2j}
\rho(r\vert Zq,{\bf 0})$ $(j=0,1,\ldots)$ are given by
\begin{equation} \label{4.5}
I_j = -Zq \kappa^2 \int_0^{\infty} {\rm d}r\, r^{2j+1} K_0(\kappa r)
= - Zq \left( \frac{2}{\kappa} \right)^{2j} 
\left[ \Gamma(1+j) \right]^2 ,
\end{equation}
where $\Gamma(x)$ denotes the Gamma function.
For $j=0$, the electroneutrality condition (\ref{1.20}) takes place.
For $j=1$, one arrives at the second-moment formula (\ref{1.22})
which is in full agreement with the general result (\ref{1.21})
taken in the weak-coupling limit $\Gamma\to 0$.

\renewcommand{\theequation}{5.\arabic{equation}}
\setcounter{equation}{0}

\section{The free-fermion point}
The fermionic representation of the 2D jellium simplifies 
substantially for the coupling constant $\Gamma=2$ ($\gamma=1$),
because the composite variables (\ref{2.6}) become the ordinary
anticommuting ones.
Having the fermionic action of the form 
$S=\sum_{j=0}^{N-1} \xi_j w_j \psi_j$ it is easy to show that
\begin{eqnarray} 
Z_N & = & \prod_{j=0}^{N-1} w_j , \label{5.1} \\
\langle \xi_j\psi_j \rangle & = & \frac{1}{w_j}
\qquad j=0,1,\ldots,N-1. \label{5.2}
\end{eqnarray} 
In the limit of the infinite disc radius $R\to\infty$, the interaction
strength (\ref{3.4}) at $\gamma=1$ is given by
\begin{equation} \label{5.3}
w_j = \frac{1}{n} \frac{1}{(\pi n)^{Z+j}} \Gamma(Z+j+1) .
\end{equation}
For an infinite number of jellium particles $N\to\infty$,
the particle density (\ref{3.5}) induced by the guest charge $Zq$
reads
\begin{equation} \label{5.4}
\frac{n(r\vert Zq,{\bf 0})}{n} = f_Z(\pi n r^2) , \qquad
f_Z(t) = {\rm e}^{-t} \sum_{j=0}^{\infty} \frac{t^{Z+j}}{\Gamma(Z+j+1)} .
\end{equation} 
It is seen that the induced density is well defined for $Z>-1$,
and this is indeed the range of the guest-charge stability (\ref{1.15}) 
for $\Gamma=2$.

Let us first treat the region of $Z>0$ 
($q>0$ will be considered for simplicity).
We shall need the incomplete Gamma function which is defined as follows
\cite{Gradshteyn}:
\begin{equation} \label{5.5}
\Gamma(Z,t) = \int_t^{\infty} {\rm d}s\, s^{Z-1} {\rm e}^{-s}
= \Gamma(Z) - \int_0^t {\rm d}s\, s^{Z-1} {\rm e}^{-s} ,
\qquad Z>0 .
\end{equation}
It can be readily shown by applying a series of integrations by parts that
\begin{equation} \label{5.6}
\Gamma(Z,t) = \Gamma(Z) - \Gamma(Z) {\rm e}^{-t}
\sum_{j=0}^{\infty} \frac{t^{Z+j}}{\Gamma(Z+j+1)} .
\end{equation}
The function $f_Z(t)$, defined in Eq. (\ref{5.4}), is therefore 
expressible as
\begin{equation} \label{5.7}
f_Z(t) = 1 - \frac{\Gamma(Z,t)}{\Gamma(Z)} 
\end{equation}
and the induced charge density reads
\begin{equation} \label{5.8}
\rho(r\vert Zq,{\bf 0}) = - q n \frac{\Gamma(Z,\pi nr^2)}{\Gamma(Z)} ,
\qquad Z>0 .
\end{equation}
Since $\partial_t \Gamma(Z,t) = - t^{Z-1} {\rm e}^{-t}$,
the derivative $\partial_r \rho(r\vert Zq,{\bf 0})$ is positive
for any value of $r$. 
Consequently, the induced charge density is the monotonically increasing 
function of $r$, going from $-qn$ at $r=0$ to $0$ at $r\to\infty$. 
The moments of the charge cloud around the guest particle
$I_j = \int_0^{\infty} 2\pi r {\rm d}r\, r^{2j} \rho(r\vert Zq,{\bf 0})$ 
$(j=0,1,\ldots)$ are given by
\begin{equation} \label{5.9}
I_j = - 2\pi q n \int_0^{\infty} {\rm d}r\, r^{2j+1}
\frac{\Gamma(Z,\pi nr^2)}{\Gamma(Z)} 
= - \frac{q}{(j+1) (\pi n)^j} \frac{\Gamma(Z+j+1)}{\Gamma(Z)} ,
\end{equation}
where we have applied an integration by parts.
For $j=0$, one recovers the electroneutrality sum rule (\ref{1.20}).
For $j=1$, one gets the result (\ref{1.23}) which is in full agreement
with the general result (\ref{1.21}) taken at $\Gamma=2$.

As concerns the stability region of negative $Z$-values $-1<Z<0$,
we first write down a recursion relation for $f_Z(t)$
following from the definition (\ref{5.4}):
\begin{equation} \label{5.10}
f_Z(t) = {\rm e}^{-t} \frac{t^Z}{\Gamma(Z+1)} + f_{Z+1}(t) .
\end{equation}
Thus,
\begin{equation} \label{5.11}
\rho(r\vert Zq,{\bf 0}) = q n \left[ {\rm e}^{-\pi nr^2}
\frac{(\pi nr^2)^Z}{\Gamma(Z+1)} -
\frac{\Gamma(Z+1,\pi nr^2)}{\Gamma(Z+1)} \right] ,
\qquad Z > -1 .
\end{equation}
The induced charge density is now the monotonically decreasing 
function of $r$, going from $\infty$ at $r=0$ to $0$ at $r\to\infty$. 
It is easy to verify that the formula for its even moments coincides
with the previous one (\ref{5.9}).
This fact permits one to extend the validity of the zeroth-moment
(\ref{1.20}) and second-moment (\ref{1.23}) sum rules to the
region of negative $Z$-values $-1<Z<0$.

\renewcommand{\theequation}{6.\arabic{equation}}
\setcounter{equation}{0}

\section{Conclusion}
In this paper, we have generalized the standard zeroth- and second-moment 
Stillinger-Lovett sum rules for the charge correlation functions to 
the presence of a guest charge immersed in the bulk interior 
of the 2D jellium.
The derivation procedure was based on the fermionic technique which
is associated specifically with the 2D jellium model.
It is an open question whether the generalization of the sum rules can be
accomplished also in higher dimensions or for many-component Coulomb fluids.
The present results might inspire specialists to establish some new
phenomenological arguments which go beyond the standard ones.

\section*{Acknowledgments}
I thank Bernard Jancovici for careful reading of the manuscript
and useful comments.
The support by grant VEGA 2/6071/26 is acknowledged.

\end{document}